\documentclass[aps,prd,reprint,superscriptaddress,twocolumn,preprintnumbers]{revtex4-2}
\usepackage{hyphenat}
\usepackage{comment}
\usepackage[utf8]{inputenc} 
\usepackage{graphicx,color,overpic,mathtools}
\usepackage{amsthm,amsmath,amssymb,hyperref,mathrsfs}
\usepackage{flexisym}
\usepackage{braket,bm,bbm,setspace}
\PassOptionsToPackage{normalem}{ulem}
\usepackage{ulem} 
\usepackage{physics}
\usepackage{float}
\usepackage[makeroom]{cancel}
\usepackage[english]{babel}
\usepackage{graphicx}
\graphicspath{ {images/} }
\usepackage{breqn}
\usepackage{bm}
\addto\captionsspanish{}
\hypersetup{
    colorlinks=false,
    pdfborder={0 0 0},
}
\usepackage[usenames,dvipsnames]{xcolor}

\makeatletter
\let\cat@comma@active\@empty
\makeatother

\begin{document}

\title{Ultracompact horizonless objects in order-reduced semiclassical gravity}

\author{Julio Arrechea}
\affiliation{IFPU, Institute for Fundamental Physics of the Universe, via Beirut 2, 34014 Trieste, Italy}
\affiliation{SISSA, International School for Advanced Studies, via Bonomea 265, 34136 Trieste, Italy}
\affiliation{INFN Sezione di Trieste,
via Valerio 2, 34127 Trieste, Italy} 
\author{Carlos Barcel\'o} 
\affiliation{Instituto de Astrof\'isica de Andaluc\'ia (IAA-CSIC),
Glorieta de la Astronom\'ia, 18008 Granada, Spain}
\author{Ra\'ul Carballo-Rubio}
\affiliation{CP3-Origins, University of Southern Denmark, Campusvej 55, DK-5230 Odense M, Denmark}
\author{Luis J. Garay} 
\affiliation{Departamento de F\'{\i}sica Te\'orica and IPARCOS, Universidad Complutense de Madrid, 28040 Madrid, Spain}

\begin{abstract}

The backreaction of quantum fields  in their vacuum state results in equilibrium structures that surpass the Buchdahl compactness limit. Such backreaction is encapsulated in the vacuum expectation value of the renormalized stress-energy tensor (RSET).
In previous works we presented analytic approximations to the RSET, obtained by  dimensional reduction, available in spherical symmetry, and showed that the  backreaction-generated solutions described ultracompact fluid spheres with a negative mass interior. Here, we derive a novel approximation to the RSET that does not rely on dimensional reduction, but rather on a reduction of the differential order. This approximation also leads to regular stars surpassing the Buchdahl limit. We conclude that this is a consequence of the negative energies associated with the Boulware vacuum which, for sufficiently compact fluid spheres, make the Misner-Sharp mass negative near the centre of spherical symmetry. Our analysis provides further cumulative evidence that quantum vacuum polarization is capable of producing new forms of stellar equilibrium with robust properties accross different analytical approximations to the RSET.
\end{abstract}

\preprint{IPARCOS-UCM-23-114}
\maketitle

\section{Introduction}
The zero-point energies of quantum fields cannot be entirely renormalized away in curved spacetimes~\cite{Parker1968,BirrellDavies1982,Wald1994,HuVerdaguer2020}. This originates vacuum-driven pressures and energy densities, simply denoted as quantum vacuum polarization hereafter.
Furthermore, during cosmological expansion and black hole formation, the non-equivalence between vacuum state definitions at early and late times manifests through the creation of particles~\cite{Parker1968,Hawking1976}. Both vacuum polarization and (part of) particle creation phenomena~\cite{Barbadoetal2011} are captured by the renormalized stress-energy tensor (RSET) of quantum fields~\cite{Wald1978,DaviesFulling1977,Fulling1977}, which contribute to spacetime curvature as described by the semiclassical Einstein equations
\begin{equation}\label{Eq:SemiEinstein}
    G^{\mu}_{~\nu}=8\pi\left(T^{\mu}_{~\nu}+\langle\hat{T}^{\mu}_{~\nu}\rangle\right).
\end{equation}
Here, $T^{\mu}_{~\nu}$ is an effectively classical stress-energy tensor (SET) and $\langle\hat{T}^{\mu}_{~\nu}\rangle$ is the renormalized expectation value in vacuum of the SET operator. As it is conventionally assumed in semiclassical analyses, we consider the classical and semiclassical SETs to be conserved independently. This causes both matter sources to affect each other only by means of their respective influence on the background spacetime, as dictated by Eqs.~\eqref{Eq:SemiEinstein}.

In truly static stellar configurations, the natural vacuum for the fields is the Boulware vacuum, which represents a physical quantum vacuum as opposed to the ``vacuum emptiness" in classical general relativity. The RSET is generally not zero even in regions in which there is no classical matter. The aim of our investigations is to illustrate how, in stellar situations where classical matter is present, the polarization of the Boulware vacuum is responsible for the existence of stars in equilibrium that are much more compact than their classical counterparts. These efforts are best framed within the ongoing search for physical mechanisms or modified gravity theories giving rise to singularity-free alternatives to black holes, some even capable of mimicking current gravitational-wave~\cite{LIGOScientific2016,LigoScientific2017,Cardoso:2019rvt} and/or very-large-baseline interferometric observations~\cite{EHT2019,EHT2022,Ayzenberg:2023hfw}.
Among the many proposals existing in the literature~(e.g., \cite{Chapline:2000en,Mazur:2001fv,MazurMottola2004,Mathur2005,HoldomRen2017,Brustein:2021lnr}), the one here presented does not require any new physics beyond quantum field theory in curved spacetimes. 

Even in spacetimes with spherical symmetry (such as the stellar spacetimes addressed here), obtaining the RSET is a computationally expensive task that requires computing a vast number of field modes with high accuracy~\cite{Andersonetal1995,Tayloretal2022}. Finding self-consistent solutions to Eqs.~\eqref{Eq:SemiEinstein} requires simultaneously computing the field modes as well as the spacetime metric which these generate and propagate onto, which is extremely complex~\cite{Medaetal2020}. In consequence, to make significant progress in understanding this semiclassical backreaction problem, it is customary to appeal to RSET approximations. One of the most common approximations in the literature is the Polyakov approximation~\cite{Parentani:1994ij,Fabbrietal2006,Carballo-Rubio:2017tlh,HoMatsuo2017,HoMatsuo2018,Arrecheaetal2020,Arrecheaetal2021,Arrechea2021b,Arrecheaetal2022}.

In a series of papers~\cite{Arrecheaetal2020,Arrecheaetal2021,Arrechea2021b,Arrecheaetal2022} we have used a regularized version of the Polyakov approximation to investigate the set of self-consistent solutions to the backreaction problem in vacuum, electrovacuum, and stellar situations. We have found, in agreement with other analyses in the literature~\cite{Hiscock1988,HoMatsuo2018,ReyesTomaselli2023,Reyes2023}, that semiclassical gravity produces non-perturbative corrections in the interior of stars approaching the Buchdahl limit $C_{R}=2m_{R}/R=8/9$, where $R$ is the radius of the star and $m_{R}=m(R)$ is the Misner-Sharp mass~\cite{MisnerSharp1964,HernandezMisner1966}, $m(r)$, evaluated on the former. The Buchdahl limit establishes the maximum compactness $C_{R}$ attainable for regular stars in equilibrium satisfying certain classically reasonable properties~\cite{Buchdahl1959,Wald:1984rg,Andreasson:2007ck,Karageorgis:2007cy,UrbanoVeermae2018,Alho:2022bki}. In particular, Buchdahl's theorem requires the total energy density to be a positive, inwards-non-decreasing function of $r$. When Eqs.~\eqref{Eq:SemiEinstein} are treated as a modified theory of gravity and solved self-consistently with a simple classical matter content (e.g., a perfect fluid with constant density), we find that the Buchdahl limit disappears. This is caused by the negative energy densities in the RSET, which grow in negativity as the compactness $C_{R}$ increases. 

Despite its great convenience, the Polyakov approximation exhibits drawbacks that hinder its use in stellar spacetimes~\cite{Arrecheaetal2022}. In this paper, we investigate this same problem but making use of a novel and completely different approximation to the RSET. We start from the Anderson-Hiscock-Samuel~\cite{Andersonetal1995} approximation for the RSET of massless scalar fields in spherical symmetry. Then, we apply an order-reduction method to obtain a second-order semiclassical system of equations. Finally, we obtain the solutions to this system of equations in the same spirit as we would do with the solutions of a modified gravity theory (see~\cite{Olmo2019} for a review of the modified gravity approach regarding stellar interiors). Remarkably, we find regular ultracompact (i.e., largely surpassing the Buchdahl limit) stellar configurations, that are qualitatively similar to those found using the regularized Polyakov method. This provides another independent evidence of the plausible absence of a Buchdahl limit in semiclassical gravity. For completeness, let us mention that this result is in tune with independent analyses of the impact of modifications of gravity on stellar structure within the asymptotic safety framework~\cite{Bonanno:2019ilz,Platania2023}. 

This paper is organized as follows. In Section~\ref{Sec:OrderReduction}, we introduce the order-reduction method for the RSET of minimally coupled fields, both in vacuum and in the presence of matter. In Section~\ref{Sec:Stars} we obtain complete numerical solutions to Eqs.~\eqref{Eq:SemiEinstein}, finding solutions describing fluid spheres surpassing the Buchdahl limit.  We also discuss their behaviour upon crossing this compactness threshold and the similarities with previous results obtained in the Polyakov approximation, which are presented in Section~\ref{Sec:Polyakov}. In Section~\ref{Sec:Buchdahl} we provide further evidence for the absence of a Buchdahl limit in semiclassical gravity, focusing on the Anderson-Hiscock-Samuel RSET prior application of the order reduction procedure.
We finish with some conclusions and further discussion in Sec.~\ref{Sec:Discussion}.

\section{Reducing the order of the AHS-RSET in the presence of matter}
\label{Sec:OrderReduction}

\subsection{The AHS-RSET and stellar spacetimes}
For simplicity, for the quantum matter sector, we will restrict our discussion to a single massless minimally coupled quantum scalar field, while the classical sector is described by a perfect fluid of constant density and isotropic pressures. We consider spherically symmetric spacetimes of the form
\begin{equation}\label{Eq:LineElement}
    ds^{2}=-f(r)dt^{2}+h(r)dr^{2}+r^{2}d\Omega^{2},
\end{equation}
where $d\Omega^{2}$ is the line element of the unit sphere, $f(r)$ is the redshift function and \mbox{$C(r)\equiv2m(r)/r=1-h(r)^{-1}$} is the compactness function and $m(r)$ the Misner-Sharp mass. 

Regarding the classical SET, we will consider an isotropic perfect fluid,
\begin{equation}\label{Eq:ClassicalSET}
    T^{\mu}_{~\nu}=\left(\rho+p\right)u^{\mu}u_{\nu}+p\delta^{\mu}_{~\nu},
\end{equation}
with $p$ and $\rho$ denoting the pressure and energy density measured by an observer comoving with the fluid with $4$-velocity $u^{\mu}$. Covariant conservation of the SET~\eqref{Eq:ClassicalSET} imposes the relation
\begin{equation}\label{Eq:Cons}\nabla_{\nu}T^{\mu}_{~r})= p'+\frac{f'}{2f}\left(\rho+p\right)=0,
\end{equation}
where primed quantities are differentiated with respect to $r$.
Additionally, we need to impose an equation of state for the fluid. For simplicity, we assume a constant density equation of state
\begin{equation}\label{Eq:EoS}
    \rho(r)\equiv \rho,
\end{equation}
which is conveniently simple and saturates the conditions of the Buchdahl compactness bound~\cite{Buchdahl1959,UrbanoVeermae2018}.

A result that is essential for the discussion below is that the exact RSET obtained in~\cite{Andersonetal1995} in spherical symmetry naturally splits into independently conserved analytic (AHS-RSET hereafter) and numeric parts: 
\begin{equation}\label{Eq:AHSRSET}
    \langle\hat{T}^{\mu}_{~\nu}\rangle_{\text{ren}}=\langle\hat{T}^{\mu}_{~\nu}\rangle_{\text{AHS}}+\langle\hat{T}^{\mu}_{~\nu}\rangle_{\text{num}}.
\end{equation}
This complete RSET is not suitable for an analytic exploration of self-consistent solutions to the semiclassical Einstein equations. However, the conserved nature of $\langle\hat{T}^{\mu}_{~\nu}\rangle_{\text{AHS}}$ makes it a suitable candidate for an analytical approximation to this problem, although with specific technical issues such as its higher-derivative nature.

One strategy to simplify the backreaction problem is to take only the analytic portion in Eq.~\eqref{Eq:AHSRSET}, which we call in the following AHS-RSET, as an approximation to the exact RSET. For a massless field, it has been shown that the analytic part is indeed a great qualitative approximation in the Schwarzschild spacetime~\cite{CandelasHoward1984,Arrecheaetal2023a} that captures the irregular behaviour of the Boulware state at event horizons~\cite{Arrecheaetal2023b}. Therefore, the elimination of the numerical part is a working hypothesis of the qualitative approximation scheme we are using; whether this numerical portion is indeed small will require its computation in stellar spacetimes, which, to the knowledge of the authors, has not been attained.

The AHS-RSET is of higher order in derivatives due to the quasi-local nature of covariant renormalization~\cite{Christiensen1976,Andersonetal1995} (explicit expressions are displayed in Appendix~\ref{Appendix:AHS}), thus making Eqs.~\eqref{Eq:SemiEinstein} a system of higher-derivative differential equations, from which it is unclear how to extract physically meaningful solutions without resorting to further approximations~\cite{Parker1968,FlanaganWald1996,Hochbergetal1997, Gaoetal2023}, in particular order reduction as explained below.

Order reduction provides a general algorithm to eliminate runaway solutions in systems with time derivatives of higher than second order, such as the Abraham-Lorentz equation~\cite{Landau1975}. In the context of semiclassical gravity, it was first used to prove the stability of Minkowski spacetime~\cite{Simon1990,FlanaganWald1996}, reinforcing the idea that not every solution to the semiclassical equations is physically meaningful. Later, the same method was applied to cosmological spacetimes~\cite{ParkerSimon1993}, and its extension to fluid spheres was also suggested. This procedure leads to the conservative viewpoint in which non-perturbative solutions are discarded as plainly non-physical, thus restricting semiclassical effects to mere perturbative corrections. Albeit this is perfectly consistent both from conceptual and formal perspectives, we are interested in modelling non-perturbative semiclassical effects that arise in the interior of stars approaching the Buchdahl limit (by non-perturbative, we mean that the RSET becomes comparable in magnitude to the classical SET, see Sec.~\ref{Sec:Buchdahl}). As discussed below, in static situations it is possible to consider the system obtained by order reduction as a self-consistent set of equations that provide an alternative version of the semiclassical Einstein equations. We will then analyze in complete detail the solutions of this system of equations, in the same spirit as in modified gravity theories.

Below we extend the perturbative order reduction method first presented in~\cite{Arrecheaetal2023}, where it was applied to the vacuum semiclassical equations, to situations where the classical SET is non-zero. We will discuss how conservation of the RSET can be reinstated after application of the order-reduction procedure by introducing adequate angular components. Applying this method to the AHS-RSET, we will construct a new RSET approximation that provides an alternative to the regularized Polyakov RSET that is still of second order in derivatives, while being however regular at $r=0$ (a characteristic absent in the standard Polyakov approximation). Despite restricting our discussion to fluids obeying the equation of state~\eqref{Eq:EoS}, this prescription can be generalized to any situation where the classical SET obeys a barotropic equation of state of the form $\rho\equiv\rho(p)$. 
The method is also valid for anisotropic perfect fluids, but we consider this extension to be out of the scope of this paper.

\subsection{The Matter-Order-Reduced RSET}
The order-reduction procedure is outlined as follows, following the same steps described in~\cite{Arrecheaetal2023}. The first step consists in taking the $tt$ and $rr$ components in the semiclassical Einstein equations~\eqref{Eq:SemiEinstein} and neglect terms of $\order{\hbar}$, such that
\begin{align}\label{Eq:Einsteinh}
    \frac{h(1-h)-rh'}{h^{2}}=
    &
    -\Omega+~\order{\hbar},\\
    \frac{rf'+f-fh}{fh}=
    &
    P+~\order{\hbar},\label{Eq:Einsteinf}
\end{align}
where the dimensionless variables \mbox{$P=8\pi r^{2}p$} and \mbox{$\Omega=8\pi r^{2}\rho$} have been chosen to simplify future expressions. 
Now, we solve Eqs.~(\ref{Eq:Einsteinh},~\ref{Eq:Einsteinf}) for $h'$ and $f'$, respectively, and differentiate them with respect to the radial coordinate $r$. 
In doing so, we derive relations between derivatives of $f,~h,~\Omega$ and $P$. For barotropic equations of state, we can specify $\Omega$ in terms of $P$, and use the conservation equation~\eqref{Eq:Cons} to further translate derivatives of $P$ into derivatives of $f$. At this stage, it is straightforward to show that Eqs.~(\ref{Eq:Einsteinh},~\ref{Eq:Einsteinf}) 
can be combined to obtain relations between derivatives of any order of $h$ and $f$ and the functions $f,~h,~\Omega$ and $P$.

Note that, in strict terms, it is only safe to neglect terms of $\order{\hbar}$ in Eqs.~(\ref{Eq:Einsteinh},~\ref{Eq:Einsteinf}) as long as the AHS-RSET is sub-leading with respect to the classical SET. That this condition is not satisfied indicates the failure of the reduction of order and, also, of the semiclassical approximation as a whole. Nonetheless, since we treat the semiclassical equations as a modified gravity, we consider the order reduced equations as just a low-differential order set of modified equations.

The procedure above can be particularized to constant density fluids, for which all derivatives of $\rho$ vanish, which results in the following relations:
\begin{align}\label{Eq:OrderRelsMatf}
    rh'=
    &
    h\left[1+\left(\Omega-1\right)h\right],\nonumber\\
    r^{2}h''=
    &
    2h^{2}\left[2\Omega-1+\left(\Omega-1\right)^{2}h\right],\nonumber\\
    r^{3}h^{(3)}=
    &
    2h^{2}\left[4\Omega+3\left(\Omega-1\right)\left(3\Omega-1\right)h+3\left(\Omega-1\right)^{3}h^{2}\right],\nonumber\\
    rf'=
    &
    -f\left[1-\left(P+1\right)h\right],\nonumber\\
    2r^{2}f''=
    &
    f\left[4+\left(\Omega+P-4\right)h+\left(\Omega+P\right)\left(P+1\right)h^{2}\right],\nonumber\\
    4r^{3}f^{(3)}=
    &
    -f\left[24-\left(\Omega+9P+24\right)h\right.\nonumber\\
    &
    \left.\quad~~~-3\left(\Omega+P\right)\left(\Omega+3P-2\right)h^{2}\right.\nonumber\\
    &
    \left.\quad~~~-3\left(\Omega+P\right)\left(\Omega-1\right)\left(P+1\right)h^{3}\right],\nonumber\\
    8r^{4}f^{(4)}=
    &
    f\left[192+\left(7\Omega-57P-192\right)h\right.\nonumber\\
    &
    \left.\quad+\left(\Omega+P\right)\left(26\Omega+63P+33\right)h^{2}\right.\nonumber\\
    &
    \left.\quad+3\left(\Omega+P\right)\left(\Omega-1\right)\left(5\Omega+22P-3\right)h^{3}\right.\nonumber\\
    &
    \left.\quad+15\left(\Omega+P\right)\left(\Omega-1\right)^{2}\left(P+1\right)h^{4}\right].
\end{align}
The expressions above can be inserted in the $\langle\hat{T}^{t}_{~t}\rangle^{\textrm{AHS}}$ and $\langle\hat{T}^{r}_{~r}\rangle^{\textrm{AHS}}$ components of the AHS-RSET~(\ref{Eq:AHSRSETtt},\ref{Eq:AHSRSETrr}), thus transforming them into quantities with no derivatives of the metric functions and hence reducing their differential order. The angular pressures of this order-reduced RSET are fixed by imposing its covariant conservation~\cite{Arrecheaetal2023}, namely
\begin{align}\label{Eq:ConsRel}
    \nabla_{\mu}\langle\hat{T}^{\mu}_{~r}\rangle=
    &
    ~\partial_{r}\langle\hat{T}^{r}_{~r}\rangle+\frac{2}{r}\left(\langle\hat{T}^{r}_{~r}\rangle-\langle\hat{T}^{\theta}_{~\theta}\rangle\right)\nonumber\\
    &
    +\frac{f'}{2f}\left(\langle\hat{T}^{r}_{~r}\rangle-\langle\hat{T}^{t}_{~t}\rangle\right)=0.
\end{align}
The resulting quantity will be denoted as the Matter-Order-Reduced RSET (or MOR-RSET, hereafter) in the rest of the paper.

Note that this procedure is not unique without further considerations. Had we inserted Eq.~\eqref{Eq:OrderRelsMatf} in the angular components~\eqref{Eq:AHSRSETthth} of the AHS-RSET as well, we would have obtained a quantity that does not satisfy~\eqref{Eq:ConsRel}. This implies it is possible to reduce the order of the $\langle\hat{T}^{t}_{~t}\rangle$ and $\langle\hat{T}^{\theta}_{~\theta}\rangle$ components, or of the $\langle\hat{T}^{r}_{~r}\rangle$ and $\langle\hat{T}^{\theta}_{~\theta}\rangle$ ones, instead of the pair we have selected above ($\langle\hat{T}^{t}_{~t}\rangle$ and $\langle\hat{T}^{r}_{~r}\rangle$). In the first case, 
specifying $\langle\hat{T}^{r}_{~r}\rangle$ would require to integrate Eq.~\eqref{Eq:ConsRel}, which is not possible in general. In the second case, the $\langle\hat{T}^{t}_{~t}\rangle$ component would end up containing derivatives of the metric functions. Thus, reducing the order of the components $\langle\hat{T}^{t}_{~t}\rangle$ and $\langle\hat{T}^{r}_{~r}\rangle$ results in the lowest-order RSET which is covariantly conserved.

For simplicity, and with the purposes of establishing a comparison with the results presented in Sec.~\ref{Sec:Polyakov} using the regularized Polyakov approximation, we only include here the expressions of the MOR-RSET in the Boulware vacuum state for the minimally coupled case (i.e. \mbox{$\kappa=0,~\xi=0$} in the notation from~\cite{Andersonetal1995}). The components of the MOR-RSET (with $\hbar=1$) take the form 
\begin{align}\label{eq:morrset}
23040\pi^{2}r^{4}h^{2}\langle\hat{T}^{\mu}_{~\nu}\rangle_{\textrm {MOR}}=\mathcal{S}^{\mu}_{~\nu}+\mathcal{T}^{\mu}_{~\nu}\log \left(\lambda^{2}f\right),
\end{align}
where
$\lambda$ is a dimensionless arbitrary parameter 
that captures local ambiguities associated with the renormalization procedure, and the diagonal tensors $\mathcal{S}^{\mu}_{~\nu}$ and $\mathcal{T}^{\mu}_{~\nu}$ have components
\begin{align}\label{Eq:MORComponents}
    \mathcal{S}^{t}_{~t}=
&
249-4\left(17\Omega+63P+105\right)h\nonumber\\
&
+2\left[\Omega\left(117\Omega+188\right)+6P\left(105\Omega+148P+53\right)\right]h^{2}\nonumber\\
&
-4\left(P+1\right)^{2}\left(77\Omega+135P-3\right)h^{3}+33\left(P+1\right)^{4}h^{4},\nonumber\\
	\mathcal{T}^{t}_{~t}=
&
\left(\Omega+P\right)\left[60h-30\left(\Omega+13P+4\right)h^{2}\right.\nonumber\\
&
\left.+60\left(P+1\right)^{2}h^{3}\right],\nonumber\\
	\mathcal{S}^{r}_{~r}=
&
-75-\left(54\Omega+382 P-84\right)h\nonumber\\
&
-\left[6\Omega\left(5\Omega+6\right)+4P\left(30\Omega-93P-100\right)-46\right]h^{2}\nonumber\\
&
+2 \left(P+1\right)^2 \left(45 \Omega+57 P-22\right)h^{3}-11 \left(P+1\right)^4 h^4,\nonumber\\
	\mathcal{T}^{r}_{~r}=
&
\left(\Omega+P\right)\left[66 h -6 \left(3 \Omega+5 P +6\right) h^2\right.\nonumber\\
&
\left.-30\left(P+1\right)^2 h^3\right],\nonumber\\
\mathcal{S}^{\theta}_{~\theta}=
&
\left(fh\right)^{-1}\times\left\{\vphantom{ \left[r\left(\Omega -1\right) f'-\left(3 P-1\right)f \right]h^5} 75 r f h'\right.\nonumber\\
&
\left.-\left\{81 r f'-\left[r \left(27 \Omega+191 P -42\right)h' +75\right]f \right\}h\right.\nonumber\\
&
\left.+\left[6 r (22 \Omega +16 P+21) f'-84 f\right]h^2\right.\nonumber\\
&
\left.-\left\{r \vphantom{\left(P^{2}\right)}\left[\Omega  (45 \Omega +221)\right.\right.\right.
\nonumber\\
&
\left.\left.\left.+3 P (175 \Omega+184 P +59)+20\right] f' \vphantom{\left[\Omega^{2}\right]}\right.\right.\nonumber\\
&
 \left.\left.-\left[r (45 \Omega+57 P -22)(P+1)^2  h'\right.\right.\right.\nonumber\\
 &
 \left.\left.\left.-2\left(15 \Omega ^2+60 \Omega P  -186 P^2+23\right)\right]f\right\}h^3\right.\nonumber\\
&
 \left.-\left(P+1\right) \left\{\vphantom{\left[5r\left(P+1\right)^2\right]}r \left[45 \Omega ^2-78 \Omega\right.\right.\right.\nonumber\\
 &
\left.\left.\left.+P \left(46 \Omega-63 P -128\right) +14\right] f'\right.\right.\nonumber\\
&
\left.\left.+\left[11 r\left(P+1\right)^3  h'\right.\right.\right.\nonumber\\
&
\left.\left.\left.-4 P \left(45\Omega+57 P -11\right)-44\vphantom{\left(P\right)^{3}}\right]f \right\}h^4\right.\nonumber\\
&
\left.+11 \left(P+1\right)^3 \left[r\left(\Omega -1\right) f'-\left(3 P-1\right)f \right]h^5\right\} ,\nonumber\\
\mathcal{T}^{\theta}_{~\theta}=
&
\left(fh\right)^{-1}\times\left\{\vphantom{\left[5r\left(P+1\right)^2\right]}\left(\Omega+P\right)\left(-33 r fhh'-15r h^2 f'\right.\right.\nonumber\\
&
\left.\left.+3\left\{\vphantom{\left[5r\left(P+1\right)^2\right]}5 r \left(\Omega+7 P+2\right) f'\right.\right.\right.\nonumber\\
&
\left.\left.\left.-\left[5r\left(P+1\right)^2 h'+6 \Omega+10 P \right]f \right\}h^3\right.\right.\nonumber\\
&
\left.\left.+15 \left(P+1\right)\left[r\left(\Omega -1\right) f'-4Pf\right] h^4\right)\vphantom{\left[5r\left(P+1\right)^2\right]}\right\}.
\end{align}

The MOR-RSET is finite at the center of regular stellar spacetimes, and reduces to the OR-RSET derived in~\cite{Arrecheaetal2023} in vacuum $(\Omega=P=0)$, for which the dependence in $\lambda$ disappears. It also coincides with the AHS-RSET when evaluated in the classical constant-density solution~\eqref{Eq:SchwStar}. The MOR-RSET depends on the arbitrary parameter $\lambda$. In general, the value of the latter is unconstrained and must be fixed experimentally. However, for constant-density stars its value is univocally determined in terms of the remaining parameters.
Indeed, upon matching the exterior vacuum geometry with the surface of a constant density fluid sphere, we must impose continuity of the redshift function at the surface, which is a necessary condition for the absence of distributional components in the stress-energy tensor~\cite{Israel1966}. For constant-density stars, there is a jump in $\Omega$ from $\Omega(r>R)=0$ to $\Omega_{R}\equiv\Omega(R)=8\pi R^{2}\rho$ at $r=R$, where $r=R$ is the star surface, which translates into a discontinuity in $\langle\hat{T}^{r}_{~r}\rangle_{\textrm{MOR}}$ at the surface via Eqs.~\eqref{Eq:SemiEinstein}. There is an analogous discontinuity in $\langle\hat{T}^{t}_{~t}\rangle_{\textrm{MOR}}$, but this translates into a jump in $h'$, already present in the classical theory and leading to no distributional sources. 
Hence, to guarantee that $f'$ is continuous at $r=R$ so that the matching between interior and exterior geometries is smooth, we require
\begin{equation}\label{Eq:Jump} \langle{\hat{T}^{r}_{~r}\rangle}_{\rm MOR}\Big|_{r=R}=\langle{\hat{T}^{r}_{~r}\rangle}_{\rm OR}\Big|_{r=R},
\end{equation}
where $\rm OR$ stands for Order-Reduced, and  $\langle{\hat{T}^{r}_{~r}\rangle}_{\rm OR}$ follows from taking $\Omega=P=0$ in the components~\eqref{Eq:MORComponents}.
Eq.~\eqref{Eq:Jump} is satisfied for the following choice of renormalization parameter
\begin{equation}\label{Eq:NuValue}
    \log \left[\lambda^{2}f(R)\right] = \frac{h(R)\left[15h(R)-5\Omega_{R}-6\right]-9}{h(R)\left[5h(R)+3\Omega_{R}+6\right]-11}.
\end{equation}
The quantities $h(R)$, $f(R)$ and $\Omega_{R}$ are the boundary conditions which, together with $P(R)=0$, specify a unique interior solution. The values of $h(R)$ and $f(R)$ are obtained by integrating the vacuum semiclassical equations with the OR-RSET from radial infinity inwards for some positive Arnowitt-Deser-Misner (ADM) mass $M$. We refer the reader to~\cite{Arrecheaetal2023} for details on this exterior solution, and restrict our discussion here to interior solutions only.

As long as the classical SET satisfies a barotropic equation of state, we can apply the order-reduction algorithm to the AHS-RSET to obtain a tensor whose radial component contains no derivatives.  Hence, since it is reasonable to expect the classical energy density $\rho$ to be discontinuous at the surface (and, consequently, $\Omega$), the renormalization parameter $\lambda$ must always be fixed through a relation similar to~\eqref{Eq:NuValue} to ensure a smooth matching between interior and exterior spacetimes. The parameter space for the case of minimal coupling and constant density is comprised by the ADM mass $M$, the star radius $R$, and the parameter $\Omega_{R}$.

\section{Stellar solutions}
\label{Sec:Stars}

\subsection{Numerical integrations}

Having derived an analytical RSET which can be implemented to study backreaction in stellar spacetimes, we proceed to integrate the order-reduced semiclassical equations, searching for regular stars that surpass the Buchdahl limit. 

We numerically integrate the semiclassical equations with the MOR-RSET in Eq.~\eqref{eq:morrset} as the source, imposing boundary conditions in the asymptotically flat region. We take a positive ADM mass $M$, which we fix to $M=5$ in the numerical solutions here presented, and integrate inwards. Note that, due to backreaction, the Misner-Sharp mass $m(r)$ is no longer constant in the exterior spacetime. In particular, we always find $m_{\rm R}>M$, indicating that the OR-RSET adds a negative mass contribution outside the star.
Next, we take a fluid sphere of surface radius $R$, which specifies its compactness $C_{\rm R}=2m_{\rm R}/R$, and obtain $\lambda$ via Eq.~\eqref{Eq:NuValue}. The only parameter left to fix is $\Omega_{R}$ (or, equivalently, $\rho$), whose value determines the regular or singular character of the interior solution. We vary $\Omega_{R}$ through several orders of magnitude seeking for solutions that are regular up to the center of the star. In practice, since the point $r=0$ is numerically unstable, we match the numerical solution with an analytic solution found from expanding the semiclassical equations in the small $r$ limit. We joint these two at a radius at least five orders of magnitude smaller than $R$, finding that the metric functions and curvature invariants are everywhere bounded.

The set of semiclassical stellar solutions is involved, containing both regular and singular solutions (see~\cite{Arrechea2021b} for an exhaustive analysis of the corresponding solutions using the regularized Polyakov approximation). Despite our subject of study being the regular subset of solutions, let us briefly discuss some distinct features of the complete solution set. We find three regimes of solutions depending on whether $\Omega_{R}$ is below or above a critical value $\Omega_{R,\text{crit}}$:
\begin{enumerate}
    \item Sub-critical solutions with $\Omega_{R}<\Omega_{R,\text{crit}}$. These solutions exhibit the same naked curvature singularity present in the exterior vacuum solution~\cite{Arrecheaetal2023a} with $\Omega=0$ (which is just a particular case within this regime). At the singularity, pressure can either diverge towards positive infinity or remain bounded, depending on the value of $\Omega_{R}$. By increasing $\Omega_{R}$, the radius where this naked singularity appears is pushed towards smaller $r$.
    \item Critical solutions with $\Omega_{R}=\Omega_{R,\text{crit}}$. This solution corresponds to the star with the lowest value of $\Omega_{R}$ that is regular. It also exhibits the largest classical pressures at $r=0$, where $p$ has a global maximum. The numerical integrations depicted in Subsec.~\ref{Subsec:Properties} correspond to critical solutions.
    \item Super-critical solutions with $\Omega_{R}>\Omega_{R,\text{crit}}$. In this regime we also find regular solutions, but with a classical pressure at $r=0$ that has a local minimum. Solutions with $\Omega_{R}\gg\Omega_{R,\text{crit}}$ can even display negative classical pressures at the center.
\end{enumerate} 

From here on, we focus just on the critical solutions, i.e., those with $\Omega_{R}=\Omega_{R,\text{crit}}$. Below the classical Buchdahl limit $C_{R}=8/9$, we obtain regular stars  whose total energy density (the sum of classical and semiclassical contributions) is nearly constant. The surface density $\Omega_{R,\text{crit}}$ receives a small and positive correction over the classical value \mbox{$\Omega_{R,\text{crit}}^{\,\text{clas}}=3C_{R}$}.
In other words, by comparing a classical constant-density star with its semiclassical counterpart such that \mbox{$\Omega_{R,\text{crit}}=\Omega_{R,\text{crit}}^{\,\text{clas}}$}, their total Misner-Sharp masses satisfy $m_{R}<m_{R}^{\text{clas}}$, this difference being related to the magnitude and sign of the RSET inside the semiclassical solution. Since, for stars well-below the classical Buchdahl limit, the RSET amounts to a perturbative correction (i.e. there is no scale that compensates for its suppression of the order of the Planck scale) the correction to $\Omega_{R,\text{crit}}^{\,\text{clas}}$ reflects this perturbative character as well. In previous works using the regularized Polyakov RSET~\cite{Arrecheaetal2022}, this implied that, on average, the semiclassical energy density is contributing negatively to the total energy density of the star. A similar result applies here, as we discuss next.

\subsection{Physical properties}
\label{Subsec:Properties}
\begin{figure*}
    \centering
    \includegraphics[width=\textwidth]{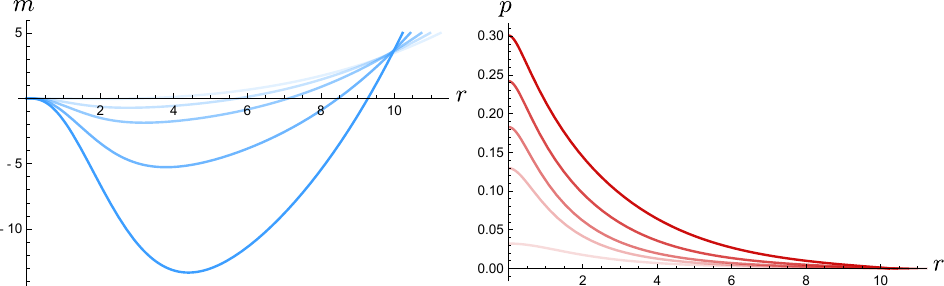}
    \caption{Misner-Sharp mass (left panel) and classical pressure (right panel) of semiclassical stars sourced by the MOR-RSET with ADM mass $M=5$ and, from lighter to darker shades, \mbox{$(C_{R},\Omega_{R},\rho)\in\{(0.89,2.720,0.008),(0.91,3.18,0.0011),(0.93,3.92,0.0013),(0.96,6.15,0.0022),(0.98,11.57,0.0044)\}$}. 
    The Misner-Sharp mass becomes more negative in the interior while central pressures grow as $C_{R}$ increases. The similarities with Fig.~\ref{Fig:MassPressPol}, which summarizes the results obtained previously in the regularized Polyakov approximation, are manifest. 
    }
\label{Fig:MassPressAHS}
\end{figure*}

Figure~\ref{Fig:MassPressAHS} shows the Misner-Sharp mass and classical pressure of a family of stars surpassing the Buchdahl limit. The Misner-Sharp mass includes contributions from the MOR-RSET that make it reach negative interior values. The region of negative mass increases in width and depth as compactness increases, but is always surrounded by an exterior layer of positive mass where the classical energy density dominates, so the total mass of the star is always positive. The pressure grows monotonically from the surface to the center, where the darker curves correspond to greater values of $C_{R}$. These large (yet finite) central pressures translate into small values of the redshift function $f$, according to the conservation equation~\eqref{Eq:Cons}. 

\begin{figure*}
    \centering
    \includegraphics[width=\textwidth]{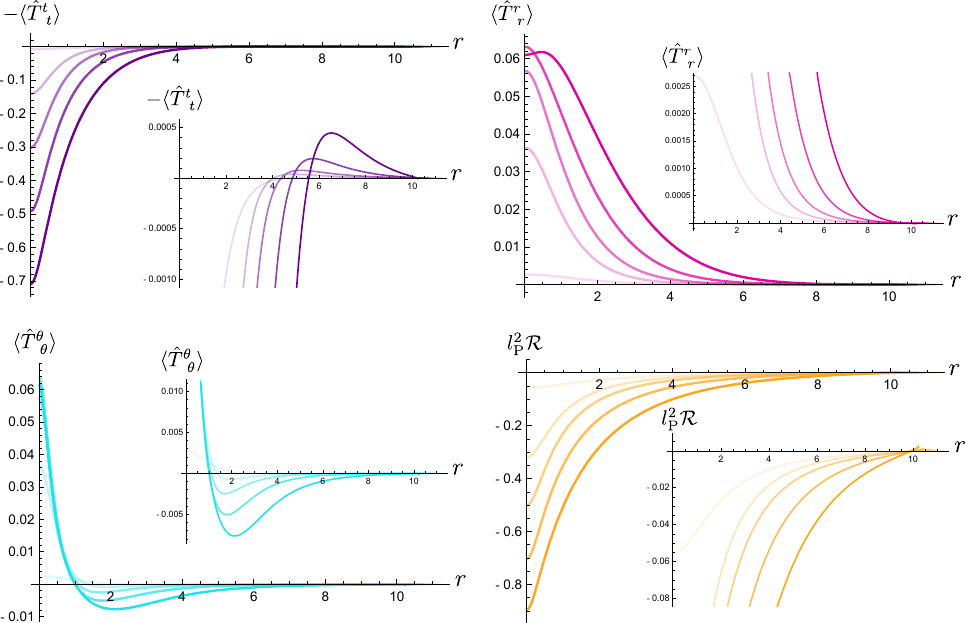}
    \caption{Numerical solutions of stars with ADM mass $M=5$ and, from lighter to darker shades, \mbox{$(C_{R},\Omega_{R},\rho)\in\{(0.89,2.720,0.008),(0.91,3.18,0.0011),(0.93,3.92,0.0013),(0.96,6.15,0.0022),(0.98,11.57,0.0044)\}$}.
    Top left, top right, bottom left, and bottom right panels display the semiclassical energy density, radial pressure, tangential pressure, and the Ricci scalar, respectively. The energy density and Ricci scalar $\mathcal{R}$ (multiplied by $l_{\rm P}^{2}=1/16\pi$) are large and negative in the interior, which is essential to allow these structures to surpass the Buchdahl limit. The semiclassical radial pressure is positive everywhere, while tangential pressures are negative in the bulk and positive at the center.}
    \label{Fig:RSETs}
\end{figure*}

Figure~\ref{Fig:RSETs} shows the temporal, radial, and angular components of the MOR-RSET. Their magnitudes visibly grow as the compactness increases. The top-left panel of Fig.~\ref{Fig:RSETs} shows the semiclassical energy density, which is positive near the surface and large and negative in the central region. Hence, in the innermost regions of the star, the total energy density is negative, thus violating the condition that the energy density must be non-decreasing inwards, which is necessary for the Buchdahl limit to hold~\cite{Buchdahl1959,Wald:1984rg,Andreasson:2007ck,Karageorgis:2007cy}. The top-right panel in Fig.~\ref{Fig:RSETs} shows the semiclassical radial pressure, which is positive and maximal at the center of the star, behaving in a similar way as the classical pressure but being slightly smaller in magnitude. The angular pressures, shown in the bottom-left panel in Fig.~\ref{Fig:RSETs}, are negative everywhere except near the center, where they change sign to match the values of the radial pressure at $r=0$, as required by regularity. From Figs.~\ref{Fig:MassPressAHS} and~\ref{Fig:RSETs} it is straightforward to reconstruct the full SET. It can be checked that the violations of the energy conditions present in the total SET are coming from the MOR-RSET, and that the classical SET satisfies all of them.

\begin{figure}
    \centering
    \includegraphics[width=\columnwidth]{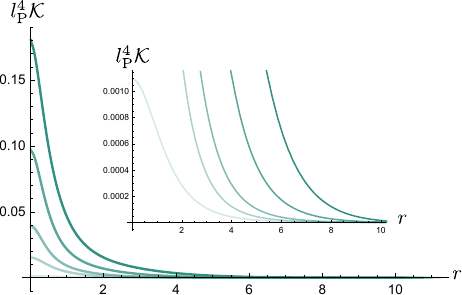}
    \caption{Plot of the Kretschmann scalar (multiplied by $l_{\rm P}^{4}=1/256\pi^{2}$ of stars with ADM mass $M=5$ and, from lighter to darker shades, $(C_{R},\Omega_{R},\rho)\in\{(0.89,2.720,0.008),(0.91,3.18,0.0011),(0.93,3.92,0.0013),\\(0.96,6.15,0.0022),(0.98,11.57,0.0044)\}$. The Kretschmann invariant remains bounded and below Planckian values throughout the whole stellar interior, being maximal at $r=0$.}
    \label{Fig:Kretschmann}
\end{figure}

The bottom-right panel in Fig.~\ref{Fig:RSETs} shows the Ricci scalar
\begin{align}\label{Eq:Rscalar}
    \mathcal{R}=
    &
    \frac{2}{r^2}\left(1-\frac{1}{h}\right)+\frac{2}{hr}\left(\frac{h'}{h}-\frac{f'}{f}+\frac{rf'h'}{4fh}\right)\nonumber\\
    &
    +\frac{1}{2h}\left[\left(\frac{f'}{f}\right)^{2}-\frac{2f''}{f}\right],
\end{align}
which is negative and has near-Planckian values. The Kretschmann scalar,
\begin{align}
 \mathcal{K}=
 &
 R_{\mu\nu\rho\sigma}R^{\mu\nu\rho\sigma}=
 \frac{4}{r^{4}}\left(1-\frac{1}{h}\right)^{2}\nonumber\\
 &
 +\frac{2}{h^{2}r^{2}}\left[\left(\frac{h'}{h}\right)^{2}+\left(\frac{f'}{f}\right)^{2}\right]\nonumber\\
 &
 +\frac{1}{4f^{4}h^{4}}\left\{f f'h'+h\left[\left(f'\right)^{2}-2ff''\right]\right\}^{2},
\end{align}
which is a positive definite curvature invariant, is shown in Fig.~\ref{Fig:Kretschmann}. Both the Ricci and Kretschmann invariants have been rescaled by the Planck length to show that they stay well below Planckian values.
Let us stress that we are not considering a large separation of scales between $\hbar$ and $R$ from a numerical perspective; we checked that the main aspects discussed above do not change when considering a broader separation of scales, in particular that the RSET regularizes the curvature singularity (e.g. divergent Ricci and Kretschmann scalars) that would be present at $r=0$ for classical stars at the Buchdahl limit.

Let us make clear that, due to the limitations of the AHS-RSET to perform self-consistent analysis, strictly speaking we cannot guarantee that our results using the MOR-RSET would be reproduced using the AHS-RSET (in any case the comparison would be only qualitative). Nonetheless, again our attitude here towards semiclassical theories of gravity is heuristic: 
we use semiclassical ideas to guide the construction of modified theories of gravity and then analyze their content without imposing constraints coming from their possible embedding in more complete frameworks (which is the same logic generally applied to the Einstein field equations). The main motivation behind this approach is to understand whether there are robust features in the space of solutions across different approximations. From this perspective, the results presented here, together with previous works, provide a clear indication of the plausibility that stellar structures 
could largely surpass the Buchdahl compactness limit. In all cases, a similar mechanism arises, which generates an effective negative energy core to support the beyond-Buchdahl stellar structure. 

\begin{figure}
    \centering
    \includegraphics{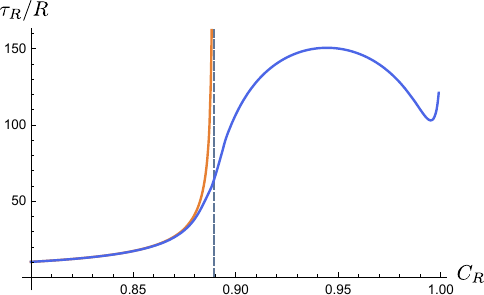}
    \caption{Plot of the crossing time needed for a null ray to travel from the surface of a star to its center and back, in terms of the surface compactness. The vertical dashed line indicates the Buchdahl limit $C_{R}=8/9$. The orange and blue curves are the crossing times of the classical and semiclassical constant-density solutions, respectively. The crossing time of the former diverges in the Buchdahl limit, while that of the latter remains finite.}
    \label{Fig:CrossingTime}
\end{figure}

The metric functions obtained in this paper can be used to calculate a number of quantities. As an example, the crossing time that a null ray emitted from the surface needs to be reflected at $r=0$ and reach $r=R$ is
\begin{equation}\label{Eq:CrossingTime}
    \tau_{R}=2 \int_{0}^{R}\left(h/f\right)^{1/2}dr'.
\end{equation}
Fig.~\ref{Fig:CrossingTime} shows how this crossing time for classical and semiclassical stars scales with $C_{R}$. We observe that the crossing time for semiclassical stars is finite upon crossing the Buchdahl threshold, contrary to what happens in the classical solution. Due to the characteristics of the exterior solution (whose details are in~\cite{Arrecheaetal2023}), the crossing time stays finite in the $C_{\rm R}\to1$ limit. Hence, these objects are appealing from a phenomenological point of view due to their capability of mimicking black holes, while also leading to potentially observable signatures such as gravitational-wave echoes~\cite{Cardoso:2016rao,CardosoPani2017} or additional rings in shadows~\cite{Carballo-Rubio:2022aed,Eichhorn:2022fcl}.

\section{Stellar equilibrium in the regularized Polyakov approximation}
\label{Sec:Polyakov}

Since estimating the backreaction effects of quantum vacuum polarization in an exact way proves to be an intractable problem, we have followed several simplifications to obtain an analytic RSET of low derivative order that allows to find self-consistent solutions to the semiclassical equations. To show whether our simplified model is adequately capturing the main physical ingredients, we will show how an entirely unrelated RSET approximation, the regularized Polyakov RSET, leads to qualitatively similar conclusions.  

The regularized Polyakov approximation is obtained when ignoring angular fluctuations of quantum fields propagating on a spherically symmetric geometry, implementing a dimensional reduction 
to an effectively 2D geometry. 
Moreover, the wave equation for the $s$-wave component of massless fields acquires, near the Schwarzschild radius, the form of the $2$D wave equation~\cite{FabbriNavarro-Salas2005}, which is conformally invariant by construction. In $2$D, the wave equation admits analytic solutions and point-splitting renormalization is much simpler than in $4$D, resulting in an analytic RSET~\cite{DaviesFulling1977} of second order in derivatives, both features being absent in the $4$D counterpart~\eqref{Eq:AHSRSET}. This $2$D RSET can be used to construct a $4$D quantity, the (regularized) Polyakov RSET, upon multiplying it by a free radial function and imposing covariant conservation, resulting in the following components:
\begin{align}\label{Eq:PolyakovRSET}
    \langle\hat{T}^{t}_{~t}\rangle^{\rm P}=
    &
    \frac{F}{96\pi h}\left[\frac{2f'h'}{fh}+3\left(\frac{f'}{f}\right)^{2}-\frac{4f''}{f}\right],\nonumber\\
    \langle\hat{T}^{r}_{~r}\rangle^{\rm P}=
    &
    -\frac{F}{96\pi h}\left(\frac{f'}{f}\right)^{2},\nonumber\\
    \langle\hat{T}^{\theta}_{~\theta}\rangle^{\rm P}=
    &
    -\frac{\left(2F+rF'\right)}{192\pi h}\left(\frac{f'}{f}\right)^{2},
\end{align}
where $F=F(r)$ is a regularizing function that must be introduced to avoid a singularity at $r=0$ coming from dimensional reduction, the specific functional form of which remains to be fixed.

\begin{figure*}
    \centering\includegraphics[width=\textwidth]{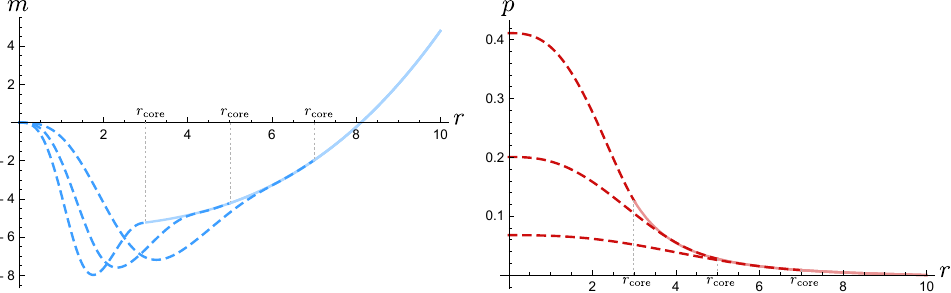}
    \caption{Misner-Sharp mass (left panel) and classical pressure (right panel) of stars obtained in the regularized Polyakov approximation. These solutions have $C_{R}=0.96$, $R=10$ and $\rho\simeq0.0025$. Continuous lines correspond to the solution with $F=1/r^{2}$, while the dashed curves represent the region where $F=F_{\text{reg}}$ for different $r_{\text{core}}$ values.
    }
\label{Fig:MassPressPol}
\end{figure*}

Within the regularized Polyakov approximation we found families of radial functions $F$ that generate solutions to Eq.~\eqref{Eq:SemiEinstein} describing regular stellar configurations surpassing the Buchdahl compactness limit~\cite{Arrecheaetal2022} with similar qualitative features as the ones obtained with the MOR-RSET and shown in Fig~\ref{Fig:MassPressAHS}. A few example solutions are displayed in Fig.~\ref{Fig:MassPressPol}, where $F$ is chosen such that
\begin{align}
    F(r)=
    &
    1/r^{2},\quad r>r_{\text{core}},\nonumber\\
    F(r)=
    &
    F_{\text{reg}}(r),\quad r\leq r_{\text{core}},
\end{align}
and $F_{\text{reg}}$ is found by reverse-engineering the semiclassical equations assuming an analytical, regular pressure ansatz (details in~\cite{Arrecheaetal2022}) and $r_{\text{core}}<R$. Regular stars surpassing the Buchdahl limit are found for any $r_{\text{core}}$, all having in common the presence of a negative mass interior with positive classical pressures. This regular, negative mass interior is possible thanks to the negative energy densities characteristic of the regularized Polyakov RSET. 

The arbitrariness in specifying the behavior of the regularized Polyakov RSET at $r=0$, due to ambiguities in $F$, motivated the consideration of other approximations that could be critically compared with the former. This provided the main motivation behind deriving the MOR-RSET which, being four-dimensional from the start, is well-defined at $r=0$. Both the regularized Polyakov RSET and the MOR-RSET have resulted in regular stars that surpass the Buchdahl limit and display similar interiors, i.e., geometries with negative Misner-Sharp mass and small values of the redshift function. These analyses serve as independent confirmations of the existence of stars surpassing the Buchdahl limit due to vacuum polarization effects.


\section{Behavior of the AHS-RSET in the Buchdahl limit}
\label{Sec:Buchdahl}

While an analysis of the space of solutions using the AHS-RSET instead of the MOR-RSET is out of the scope of this paper, we can still highlight some features of the AHS-RSET with the goal of stimulating further research. The arising of beyond-Buchdahl structures supported by the MOR-RSET stems from its growth in magnitude when the Buchdahl limit is approached. We can check whether a similar growth is displayed by the AHS-RSET when calculated in fixed stellar backgrounds on the verge of reaching the Buchdahl limit. To study this in complete generality, in this section we extend the analysis to non-minimally coupled fields. 
Let us consider the spacetime of a uniform-density star and evaluate the AHS-RSET over it. Taking $\rho=3C_{R}/8\pi R^{2}$ (a necessary condition for the regularity of the metric at $r=0$~\cite{Arrechea2021b}), the corresponding interior line element~\cite{Schwarzschild1916} is
\begin{align}\label{Eq:SchwStar}
    ds^{2}=
    &
    -\frac{1}{4}\left(3\sqrt{1-C_{R}}-\sqrt{1-r^{2}C_{R}/R^{2}}\right)^{2}dt^{2}\nonumber\\
    &
    +
    \left(1-r^{2}C_{R}/R^{2}\right)^{-1}dr^{2}
    +r^{2}d\Omega^{2}.
\end{align}
Now, focusing on a stellar solution whose compactness approaches the Buchdahl limit, i.e.,
\begin{equation}\label{Eq:Buchdahl}
   C_{R}=\frac{8}{9}-\epsilon,\quad \epsilon\to0^{+},
\end{equation}
we evaluate the AHS-RSET, whose components are available in Appendix~\ref{Appendix:AHS}, at $r=0$~\eqref{Eq:AHSRSET} in the metric~\eqref{Eq:SchwStar} and in the limit defined in Eq.~\eqref{Eq:Buchdahl}. The result is
\begin{align}\label{Eq:RSETBuchdahl}
    \langle\hat{T}^{t}_{~t}\rangle_{\rm AHS}\big|_{r=0}=
    &
    -\frac{l_{\rm P}^{2}}{R^{4}}\left(\xi-\frac{1}{6}\right)^{2}\frac{\log\epsilon}{\epsilon^{2}}+\mathcal{O}\left(\epsilon^{-2}\right),\nonumber\\
    \langle\hat{T}^{r}_{~r}\rangle_{\rm AHS}\big|_{r=0}=
    &
    \langle\hat{T}^{\theta}_{~\theta}\rangle_{\rm AHS}\big|_{r=0}=\frac{1}{3}\langle\hat{T}^{t}_{~t}\rangle_{\rm AHS}\big|_{r=0},
\end{align}
with $l_{\rm P}^{2}=16/729\pi^{2}$ and $\xi$ is the coupling of the field to the Ricci scalar. In view of the above, the semiclassical energy density and pressure diverge towards negative and positive infinity, respectively, following a $\rho=-3p$ relation. It is remarkable that this leading behavior is independent of the parameters controlling the renormalization parameter $\lambda$ and the coupling $\xi$. 

Also, it is straightforward to check that the semiclassical energy density in Eq.~\eqref{Eq:RSETBuchdahl} becomes comparable in magnitude to the classical energy density approximately when
\begin{equation}\label{Eq:epsilon}
    \epsilon\simeq\mathcal{O}\left[\left(\frac{l_{\rm P}}{R}\right)\log\left(\frac{R}{l_{\rm P}}\right)\right],\quad \text{with~} R\gg l_{\rm P}.
\end{equation}
Hence, the leading-order terms in the AHS-RSET show a universal build-up of negative energy density at the center of the star when its compactness is close to the Buchdahl limit, namely when $\epsilon$ satisfies Eq.~\eqref{Eq:epsilon}. 

Therefore, even though we have not used the AHS-RSET for the self-consistent integrations, it appears that its effect when approaching the Buchdahl limit is introducing a negative energy core in the configuration which, based on our previous discussion using the MOR-RSET (and also the regularized Polyakov approximation in previous papers), might allow the existence of ultracompact configurations.

While this observation applies to the analytical piece in Eq.~\eqref{Eq:AHSRSET}, it would also be interesting to understand the behavior of the numerical piece in the latter equation.
The numerical piece is also independent of $\lambda$ but one would need to compute it to exactly determine the leading-order divergence in the complete RSET.
Also, it is important to stress that our analysis in this section has made the reasonable assumption that the larger values of the RSET would appear at the center of the configuration. That is the place in which we are able to obtain a robust analysis of the behaviour of the RSET components.

In summary, we can conclude that stellar configurations approaching the Buchdahl limit will likely develop negative energy densities at their central regions that will be comparable in magnitude to the classical energy density itself when using the AHS-RSET. 
As happens with the MOR-RSET (and the regularized Polyakov RSET), this can lead to an entire disappearance of the Buchdahl limit when the backreaction effects of the RSET are included.

\section{Conclusions}
\label{Sec:Discussion}

The Buchdahl limit in general relativity imposes an upper bound to the compactness of fluid spheres in equilibrium, as long as their density is non-decreasing
from the surface towards the center, among other conditions. We have shown that this limit disappears when incorporating the self-consistent backreaction of the RSET of a massless, minimally coupled scalar field. To be able to solve the semiclassical backreaction problem, we have derived a novel order-reduced RSET approximation that is well defined in stellar spacetimes.

Our integrations show that the Buchdahl limit disappears due to the build-up of negative energy densities in the interior of the star. 
While the approximation constructed here presents some renormalization ambiguities controlled by a parameter $\lambda$, we have been able to remove these ambiguities for a specific equation of state and by imposing the absence of distributional SETs at the star's surface. Within the range of parameters explored in this work, we conclude that there is no trace of the Buchdahl limit once vacuum backreaction is incorporated. The astonishing similarity between the solutions discussed here and those derived through the regularized Polyakov approximation~\cite{Arrecheaetal2022} is a strong indication of the robustness of this result.

Despite the ambiguities in the renormalization procedure and the different possible approximations that can be considered to find an analytical RSET, it is clear that, if quantum vacuum polarization behaves in the way described here, ultracompact stars 
should have a negative-energy core and an external (thin) layer where mass quickly regains positive values at the surface to smoothly match the vacuum exterior.
These ultracompact stars might be formed as the final equilibrium configuration resulting from a modified collapse process with regularized singularities and evanescent trapping horizons~\cite{Barceloetal2022,Carballo-Rubio:2022nuj,Carballo-Rubio:2023mvr}.

In summary, our results here motivate further research on analytical approximations able to capture the backreaction of quantum vacuum polarization, with the aim of gaining a
better understanding of the robustness and generality of the resulting stellar solutions.

\section{Acknowledgements}
The authors thank Valentin Boyanov and \mbox{Gerardo} Garc\'{\i}a-Moreno for very useful discussions. Financial support was provided by the Spanish Government through the projects PID2020-118159GB-C43/AEI/10.13039/501100011033, PID2020-118159GB-C44/AEI/10.13039/501100011033, PID2019-107847RB-C44/AEI/10.13039/501100011033, and by the Junta de Andalucía through the project FQM219. This research is supported by a research grant (29405) from VILLUM fonden. CB, JA and RCR acknowledges financial support from the Severo Ochoa grant CEX2021-001131-S funded by MCIN/AEI/ 10.13039/501100011033. JA acknowledges funding from the Italian Ministry of Education and Scientific Research (MIUR) under the grant
PRIN MIUR 2017-MB8AEZ.

\appendix

\section{The Anderson-Hiscock-Samuel RSET}
\label{Appendix:AHS}
Below we show the components of the AHS-RSET for a massless field with arbitrary coupling $\xi$ in the Boulware state. 
\begin{widetext}
\begin{dmath*}
1440\pi^{2}\langle\hat{T}^{t}_{t}\rangle_{\rm AHS}=
\frac{7 \left(f'\right)^4}{32 f^4 h^2}+\frac{7 \left(h'\right)^3}{h^5 r}+\frac{3 \left(f'\right)^2}{4 f^2 h^2 r^2}+\frac{5 f' h'}{2 f h^3 r^2}+\frac{\left(f'\right)^2 f''}{8 f^3 h^2}+\frac{19 \left(h'\right)^2 f''}{8 f h^4}+\frac{9 f' h' f''}{8 f^2 h^3}+\frac{\left(f'\right)^2 h''}{4 f^2 h^3}+\frac{f' h''}{2 f h^3 r}+\frac{13 f' h' h''}{8 f h^4}+\frac{h''}{h^3 r^2}+\frac{2 f^{(3)}}{f h^2 r}+\frac{h^{(3)}}{h^3 r}+\frac{f^{(4)}}{2 f h^2}-\frac{f' f^{(3)}}{2 f^2 h^2}-\frac{3 \left(f''\right)^2}{8 f^2 h^2}-\frac{f'' h''}{f h^3}-\frac{3 h' f^{(3)}}{2 f h^3}-\frac{f' h^{(3)}}{4 f h^3}-\frac{\left(f'\right)^3 h'}{16 f^3 h^3}-\frac{19 \left(f'\right)^2 \left(h'\right)^2}{32 f^2 h^4}-\frac{7 f' \left(h'\right)^3}{4 f h^5}-\frac{3 \left(f'\right)^3}{4 f^3 h^2 r}-\frac{2 h' f''}{f h^3 r}-\frac{5 \left(f'\right)^2 h'}{4 f^2 h^3 r}-\frac{13 h' h''}{2 h^4 r}-\frac{3 f' \left(h'\right)^2}{2 f h^4 r}-\frac{7 \left(h'\right)^2}{4 h^4 r^2}-\frac{2 h'}{h^3 r^3}+\frac{1}{r^4}-\frac{1}{h^2 r^4}+\left(-\frac{49 \left(f'\right)^4}{32 f^4 h^2}-\frac{29 h' \left(f'\right)^3}{16 f^3 h^3}+\frac{11 \left(f'\right)^3}{8 f^3 h^2 r}+\frac{3 h' \left(f'\right)^2}{2 f^2 h^3 r}+\frac{29 f'' \left(f'\right)^2}{8 f^3 h^2}+\frac{3 h'' \left(f'\right)^2}{4 f^2 h^3}-\frac{57 \left(h'\right)^2 \left(f'\right)^2}{32 f^2 h^4}+\frac{5 \left(f'\right)^2}{8 f^2 h^2 r^2}+\frac{13 \left(h'\right)^2 f'}{8 f h^4 r}+\frac{5 h' f'}{4 f h^3 r^2}+\frac{27 h' f'' f'}{8 f^2 h^3}+\frac{13 h' h'' f'}{8 f h^4}-\frac{3 f^{(3)} f'}{2 f^2 h^2}-\frac{h^{(3)} f'}{4 f h^3}-\frac{7 \left(h'\right)^3 f'}{4 f h^5}-\frac{13 f'' f'}{4 f^2 h^2 r}-\frac{3 h'' f'}{4 f h^3 r}+\frac{7 \left(h'\right)^3}{2 h^5 r}+\frac{19 \left(h'\right)^2 f''}{8 f h^4}+\frac{h''}{2 h^3 r^2}+\frac{2 f^{(3)}}{f h^2 r}+\frac{h^{(3)}}{2 h^3 r}+\frac{f^{(4)}}{2 f h^2}-\frac{9 \left(f''\right)^2}{8 f^2 h^2}-\frac{f'' h''}{f h^3}-\frac{3 h' f^{(3)}}{2 f h^3}-\frac{13 h' f''}{4 f h^3 r}-\frac{13 h' h''}{4 h^4 r}-\frac{7 \left(h'\right)^2}{8 h^4 r^2}-\frac{h'}{h^3 r^3}+\frac{1}{2 r^4}-\frac{1}{2 h^2 r^4}\right) \log\left( \lambda^{2}f\right)\end{dmath*}
\begin{dmath}\label{Eq:AHSRSETtt}
+\left(-\frac{945 \left(f'\right)^4}{8 f^4 h^2}-\frac{495 h' \left(f'\right)^3}{4 f^3 h^3}+\frac{165 \left(f'\right)^3}{f^3 h^2 r}+\frac{405 h' \left(f'\right)^2}{2 f^2 h^3 r}+\frac{495 f'' \left(f'\right)^2}{2 f^3 h^2}+\frac{45 h'' \left(f'\right)^2}{f^2 h^3}-\frac{855 \left(h'\right)^2 \left(f'\right)^2}{8 f^2 h^4}+\frac{45 \left(f'\right)^2}{2 f^2 h r^2}-\frac{45 \left(f'\right)^2}{2 f^2 h^2 r^2}+\frac{285 \left(h'\right)^2 f'}{f h^4 r}+\frac{15 h' f'}{f h^2 r^2}+\frac{405 h' f'' f'}{2 f^2 h^3}+\frac{195 h' h'' f'}{2 f h^4}-\frac{90 f^{(3)} f'}{f^2 h^2}-\frac{15 h^{(3)} f'}{f h^3}-\frac{105 \left(h'\right)^3 f'}{f h^5}-\frac{270 f'' f'}{f^2 h^2 r}-\frac{120 h'' f'}{f h^3 r}-\frac{45 h' f'}{f h^3 r^2}+\frac{60 \left(h'\right)^2}{h^4 r^2}+\frac{90 h'}{h^3 r^3}+\frac{285 \left(h'\right)^2 f''}{2 f h^4}+\frac{30 f''}{f h^2 r^2}+\frac{195 h' h''}{h^4 r}+\frac{120 f^{(3)}}{f h^2 r}+\frac{30 f^{(4)}}{f h^2}-\frac{135 \left(f''\right)^2}{2 f^2 h^2}-\frac{60 f'' h''}{f h^3}-\frac{90 h' f^{(3)}}{f h^3}-\frac{30 h^{(3)}}{h^3 r}-\frac{270 h' f''}{f h^3 r}-\frac{210 \left(h'\right)^3}{h^5 r}-\frac{30 f''}{f h r^2}-\frac{30 h''}{h^3 r^2}-\frac{30 h'}{h^2 r^3}-\frac{60}{h r^4}+\frac{60}{h^2 r^4}\right) \left(\xi -\frac{1}{6}\right)
+\left[\frac{945 \left(f'\right)^4}{4 f^4 h^2}+\frac{270 h' \left(f'\right)^3}{f^3 h^3}-\frac{720 \left(f'\right)^3}{f^3 h^2 r}+\frac{855 \left(h'\right)^2 \left(f'\right)^2}{4 f^2 h^4}-\frac{540 f'' \left(f'\right)^2}{f^3 h^2}-\frac{90 h'' \left(f'\right)^2}{f^2 h^3}-\frac{360 h' \left(f'\right)^2}{f^2 h^3 r}+\frac{270 \left(f'\right)^2}{f^2 h r^2}-\frac{90 \left(f'\right)^2}{f^2 h^2 r^2}+\frac{900 \left(h'\right)^2 f'}{f h^4 r}+\frac{180 h' f'}{f h^2 r^2}+\frac{900 f'' f'}{f^2 h^2 r}+\frac{180 f^{(3)} f'}{f^2 h^2}-\frac{405 h' f'' f'}{f^2 h^3}-\frac{360 h'' f'}{f h^3 r}-\frac{900 h' f'}{f h^3 r^2}+\frac{180 \left(h'\right)^2}{h^4 r^2}+\frac{135 \left(f''\right)^2}{f^2 h^2}+\frac{360 h'}{h^2 r^3}+\frac{360 f''}{f h^2 r^2}-\frac{360 h' f''}{f h^3 r}-\frac{360 f''}{f h r^2}-\frac{360 h'}{h^3 r^3}+\frac{180}{r^4}-\frac{360}{h r^4}+\frac{180}{h^2 r^4}+\left(-\frac{2205 \left(f'\right)^4}{8 f^4 h^2}-\frac{1305 h' \left(f'\right)^3}{4 f^3 h^3}+\frac{585 \left(f'\right)^3}{f^3 h^2 r}+\frac{675 h' \left(f'\right)^2}{f^2 h^3 r}+\frac{1305 f'' \left(f'\right)^2}{2 f^3 h^2}+\frac{135 h'' \left(f'\right)^2}{f^2 h^3}-\frac{2565 \left(h'\right)^2 \left(f'\right)^2}{8 f^2 h^4}-\frac{90 \left(f'\right)^2}{f^2 h^2 r^2}+\frac{630 \left(h'\right)^2 f'}{f h^4 r}+\frac{90 h' f'}{f h^3 r^2}+\frac{1215 h' f'' f'}{2 f^2 h^3}+\frac{585 h' h'' f'}{2 f h^4}-\frac{270 f^{(3)} f'}{f^2 h^2}-\frac{45 h^{(3)} f'}{f h^3}-\frac{315 \left(h'\right)^3 f'}{f h^5}-\frac{990 f'' f'}{f^2 h^2 r}-\frac{270 h'' f'}{f h^3 r}+\frac{450 \left(h'\right)^2}{h^4 r^2}+\frac{360 h'}{h^3 r^3}+\frac{855 \left(h'\right)^2 f''}{2 f h^4}+\frac{1170 h' h''}{h^4 r}+\frac{360 f^{(3)}}{f h^2 r}+\frac{90 f^{(4)}}{f h^2}-\frac{405 \left(f''\right)^2}{2 f^2 h^2}-\frac{180 f'' h''}{f h^3}-\frac{270 h' f^{(3)}}{f h^3}-\frac{180 h^{(3)}}{h^3 r}-\frac{720 h' f''}{f h^3 r}-\frac{1260 \left(h'\right)^3}{h^5 r}-\frac{180 h''}{h^3 r^2}+\frac{90}{r^4}-\frac{540}{h r^4}+\frac{450}{h^2 r^4}\right) \log\left(\lambda^{2} f\right)\right] \left(\xi -\frac{1}{6}\right)^2,
\end{dmath}
\begin{dmath}\label{Eq:AHSRSETrr}
1440\pi^{2}\langle\hat{T}^{r}_{r}\rangle_{\rm AHS}=\frac{\left(f'\right)^4}{32 f^4 h^2}+\frac{7 \left(f'\right)^2 \left(h'\right)^2}{32 f^2 h^4}+\frac{2 f'}{f h^2 r^3}+\frac{\left(f'\right)^3 h'}{16 f^3 h^3}+\frac{f' h'}{2 f h^3 r^2}+\frac{2 f' f''}{f^2 h^2 r}+\frac{h' f''}{f h^3 r}+\frac{f' h''}{f h^3 r}+\frac{f' f^{(3)}}{4 f^2 h^2}-\frac{\left(f''\right)^2}{8 f^2 h^2}-\frac{\left(f'\right)^2 f''}{8 f^3 h^2}-\frac{f' h' f''}{4 f^2 h^3}-\frac{\left(f'\right)^2 h''}{8 f^2 h^3}-\frac{f^{(3)}}{f h^2 r}-\frac{\left(f'\right)^3}{2 f^3 h^2 r}-\frac{\left(f'\right)^2 h'}{4 f^2 h^3 r}-\frac{7 f' \left(h'\right)^2}{4 f h^4 r}-\frac{2 f''}{f h^2 r^2}+\left(\frac{7 \left(f'\right)^4}{32 f^4 h^2}+\frac{3 h' \left(f'\right)^3}{16 f^3 h^3}-\frac{5 \left(f'\right)^3}{8 f^3 h^2 r}+\frac{7 \left(h'\right)^2 \left(f'\right)^2}{32 f^2 h^4}-\frac{3 f'' \left(f'\right)^2}{8 f^3 h^2}-\frac{h'' \left(f'\right)^2}{8 f^2 h^3}-\frac{h' \left(f'\right)^2}{2 f^2 h^3 r}-\frac{\left(f'\right)^2}{8 f^2 h^2 r^2}+\frac{h' f'}{4 f h^3 r^2}+\frac{3 f'' f'}{2 f^2 h^2 r}+\frac{h'' f'}{2 f h^3 r}+\frac{f^{(3)} f'}{4 f^2 h^2}-\frac{h' f'' f'}{4 f^2 h^3}-\frac{7 \left(h'\right)^2 f'}{8 f h^4 r}+\frac{f'}{f h^2 r^3}+\frac{7 \left(h'\right)^2}{8 h^4 r^2}+\frac{h' f''}{2 f h^3 r}-\frac{\left(f''\right)^2}{8 f^2 h^2}-\frac{f^{(3)}}{2 f h^2 r}-\frac{f''}{f h^2 r^2}-\frac{h''}{2 h^3 r^2}+\frac{1}{2 r^4}-\frac{1}{2 h^2 r^4}\right) \log \left(\lambda^{2}f\right)
+\left(\frac{135 \left(f'\right)^4}{8 f^4 h^2}+\frac{45 h' \left(f'\right)^3}{4 f^3 h^3}+\frac{45 \left(f'\right)^3}{f^3 h^2 r}+\frac{105 \left(h'\right)^2 \left(f'\right)^2}{8 f^2 h^4}-\frac{45 f'' \left(f'\right)^2}{2 f^3 h^2}-\frac{15 h'' \left(f'\right)^2}{2 f^2 h^3}-\frac{15 \left(f'\right)^2}{2 f^2 h r^2}-\frac{165 \left(f'\right)^2}{2 f^2 h^2 r^2}+\frac{105 \left(h'\right)^2 f'}{2 f h^4 r}+\frac{15 f^{(3)} f'}{f^2 h^2}-\frac{15 h' f'' f'}{f^2 h^3}-\frac{75 f'' f'}{f^2 h^2 r}-\frac{30 h'' f'}{f h^3 r}-\frac{30 h' f'}{f h^3 r^2}+\frac{30 f'}{f h r^3}-\frac{90 f'}{f h^2 r^3}+\frac{60 f''}{f h^2 r^2}+\frac{30 f^{(3)}}{f h^2 r}-\frac{15 \left(f''\right)^2}{2 f^2 h^2}-\frac{30 h' f''}{f h^3 r}\right) \left(\xi -\frac{1}{6}\right)+\left[-\frac{45 \left(f'\right)^4}{2 f^4 h^2}-\frac{45 h' \left(f'\right)^3}{2 f^3 h^3}+\frac{45 f'' \left(f'\right)^2}{f^3 h^2}-\frac{180 h' \left(f'\right)^2}{f^2 h^3 r}-\frac{90 \left(f'\right)^2}{f^2 h r^2}+\frac{450 \left(f'\right)^2}{f^2 h^2 r^2}+\frac{180 f'' f'}{f^2 h^2 r}-\frac{360 h' f'}{f h^3 r^2}-\frac{360 f'}{f h r^3}+\frac{360 f'}{f h^2 r^3}+\left(\frac{315 \left(f'\right)^4}{8 f^4 h^2}+\frac{135 h' \left(f'\right)^3}{4 f^3 h^3}+\frac{90 \left(f'\right)^3}{f^3 h^2 r}+\frac{315 \left(h'\right)^2 \left(f'\right)^2}{8 f^2 h^4}+\frac{45 h' \left(f'\right)^2}{f^2 h^3 r}-\frac{135 f'' \left(f'\right)^2}{2 f^3 h^2}-\frac{45 h'' \left(f'\right)^2}{2 f^2 h^3}-\frac{360 \left(f'\right)^2}{f^2 h^2 r^2}+\frac{315 \left(h'\right)^2 f'}{f h^4 r}+\frac{45 f^{(3)} f'}{f^2 h^2}-\frac{45 h' f'' f'}{f^2 h^3}-\frac{270 f'' f'}{f^2 h^2 r}-\frac{180 h'' f'}{f h^3 r}-\frac{360 h' f'}{f h^3 r^2}-\frac{360 f'}{f h^2 r^3}+\frac{630 \left(h'\right)^2}{h^4 r^2}+\frac{360 f''}{f h^2 r^2}+\frac{180 f^{(3)}}{f h^2 r}-\frac{45 \left(f''\right)^2}{2 f^2 h^2}-\frac{180 h' f''}{f h^3 r}-\frac{360 h''}{h^3 r^2}+\frac{90}{r^4}+\frac{540}{h r^4}-\frac{630}{h^2 r^4}\right) \log \left(\lambda^{2}f\right)\right] \left(\xi -\frac{1}{6}\right)^2,
\end{dmath}
\begin{dmath*}
1440\pi^{2}\langle\hat{T}^{\theta}_{\theta}\rangle_{\rm AHS}=\frac{17 \left(f'\right)^4}{32 f^4 h^2}+\frac{7 f' \left(h'\right)^3}{4 f h^5}+\frac{31 \left(f'\right)^2 \left(h'\right)^2}{32 f^2 h^4}+\frac{7 \left(f''\right)^2}{8 f^2 h^2}+\frac{13 \left(f'\right)^3 h'}{16 f^3 h^3}+\frac{f' f''}{f^2 h^2 r}+\frac{11 h' f''}{4 f h^3 r}+\frac{f''}{f h^2 r^2}+\frac{f' h''}{4 f h^3 r}+\frac{f'' h''}{f h^3}+\frac{3 f' f^{(3)}}{4 f^2 h^2}+\frac{3 h' f^{(3)}}{2 f h^3}+\frac{f' h^{(3)}}{4 f h^3}-\frac{f^{(4)}}{2 f h^2}-\frac{13 \left(f'\right)^2 f''}{8 f^3 h^2}-\frac{2 f' h' f''}{f^2 h^3}-\frac{3 \left(f'\right)^2 h''}{8 f^2 h^3}-\frac{19 \left(h'\right)^2 f''}{8 f h^4}-\frac{13 f' h' h''}{8 f h^4}-\frac{3 f^{(3)}}{2 f h^2 r}-\frac{\left(f'\right)^3}{2 f^3 h^2 r}-\frac{3 \left(f'\right)^2 h'}{4 f^2 h^3 r}-\frac{3 f' \left(h'\right)^2}{4 f h^4 r}-\frac{3 f' h'}{2 f h^3 r^2}-\frac{f'}{f h^2 r^3}+\left(\frac{21 \left(f'\right)^4}{32 f^4 h^2}+\frac{13 h' \left(f'\right)^3}{16 f^3 h^3}-\frac{3 \left(f'\right)^3}{8 f^3 h^2 r}+\frac{25 \left(h'\right)^2 \left(f'\right)^2}{32 f^2 h^4}-\frac{13 f'' \left(f'\right)^2}{8 f^3 h^2}-\frac{5 h'' \left(f'\right)^2}{16 f^2 h^3}-\frac{h' \left(f'\right)^2}{2 f^2 h^3 r}-\frac{\left(f'\right)^2}{4 f^2 h^2 r^2}+\frac{7 \left(h'\right)^3 f'}{8 f h^5}+\frac{7 f'' f'}{8 f^2 h^2 r}+\frac{h'' f'}{8 f h^3 r}+\frac{5 f^{(3)} f'}{8 f^2 h^2}+\frac{h^{(3)} f'}{8 f h^3}-\frac{25 h' f'' f'}{16 f^2 h^3}-\frac{13 h' h'' f'}{16 f h^4}-\frac{3 \left(h'\right)^2 f'}{8 f h^4 r}-\frac{3 h' f'}{4 f h^3 r^2}-\frac{f'}{2 f h^2 r^3}+\frac{5 \left(f''\right)^2}{8 f^2 h^2}+\frac{h'}{2 h^3 r^3}+\frac{11 h' f''}{8 f h^3 r}+\frac{f''}{2 f h^2 r^2}+\frac{13 h' h''}{8 h^4 r}+\frac{f'' h''}{2 f h^3}+\frac{3 h' f^{(3)}}{4 f h^3}-\frac{f^{(4)}}{4 f h^2}-\frac{19 \left(h'\right)^2 f''}{16 f h^4}-\frac{3 f^{(3)}}{4 f h^2 r}-\frac{h^{(3)}}{4 h^3 r}-\frac{7 \left(h'\right)^3}{4 h^5 r}-\frac{1}{2 r^4}+\frac{1}{2 h^2 r^4}\right) \log \left(\lambda^{2}f\right)+\left(-\frac{645 \left(f'\right)^4}{8 f^4 h^2}-\frac{675 h' \left(f'\right)^3}{8 f^3 h^3}+\frac{90 \left(f'\right)^3}{f^3 h^2 r}+\frac{405 h' \left(f'\right)^2}{4 f^2 h^3 r}+\frac{675 f'' \left(f'\right)^2}{4 f^3 h^2}+\frac{30 h'' \left(f'\right)^2}{f^2 h^3}-\frac{285 \left(h'\right)^2 \left(f'\right)^2}{4 f^2 h^4}-\frac{15 \left(f'\right)^2}{2 f^2 h r^2}+\frac{45 \left(f'\right)^2}{2 f^2 h^2 r^2}+\frac{225 \left(h'\right)^2 f'}{4 f h^4 r}+\frac{135 h' f'}{2 f h^3 r^2}+\frac{135 h' f'' f'}{f^2 h^3}+\frac{195 h' h'' f'}{4 f h^4}-\frac{60 f^{(3)} f'}{f^2 h^2}-\frac{15 h^{(3)} f'}{2 f h^3}-\frac{105 \left(h'\right)^3 f'}{2 f h^5}-\frac{285 f'' f'}{2 f^2 h^2 r}-\frac{45 h'' f'}{2 f h^3 r}-\frac{15 h' f'}{2 f h^2 r^2}+\frac{30 f'}{f h^2 r^3}+\frac{285 \left(h'\right)^2 f''}{4 f h^4}+\frac{15 f''}{f h r^2}+\frac{45 f^{(3)}}{f h^2 r}+\frac{15 f^{(4)}}{f h^2}-\frac{45 \left(f''\right)^2}{f^2 h^2}-\frac{30 f'' h''}{f h^3}-\frac{45 h' f^{(3)}}{f h^3}-\frac{90 h' f''}{f h^3 r}-\frac{45 f''}{f h^2 r^2}\right) \left(\xi -\frac{1}{6}\right)
\end{dmath*}
\begin{dmath}\label{Eq:AHSRSETthth}
+\left[\frac{405 \left(f'\right)^4}{2 f^4 h^2}+\frac{225 h' \left(f'\right)^3}{f^3 h^3}-\frac{495 \left(f'\right)^3}{f^3 h^2 r}+\frac{405 \left(h'\right)^2 \left(f'\right)^2}{2 f^2 h^4}-\frac{450 f'' \left(f'\right)^2}{f^3 h^2}-\frac{90 h'' \left(f'\right)^2}{f^2 h^3}-\frac{405 h' \left(f'\right)^2}{f^2 h^3 r}+\frac{90 \left(f'\right)^2}{f^2 h r^2}-\frac{270 \left(f'\right)^2}{f^2 h^2 r^2}+\frac{810 \left(h'\right)^2 f'}{f h^4 r}+\frac{90 h' f'}{f h^2 r^2}+\frac{630 f'' f'}{f^2 h^2 r}+\frac{180 f^{(3)} f'}{f^2 h^2}-\frac{360 h' f'' f'}{f^2 h^3}-\frac{360 h'' f'}{f h^3 r}-\frac{270 h' f'}{f h^3 r^2}+\frac{540 f'}{f h r^3}-\frac{540 f'}{f h^2 r^3}+\frac{90 \left(f''\right)^2}{f^2 h^2}+\frac{180 f''}{f h^2 r^2}-\frac{180 h' f''}{f h^3 r}-\frac{180 f''}{f h r^2}+\left(-\frac{1755 \left(f'\right)^4}{8 f^4 h^2}-\frac{1035 h' \left(f'\right)^3}{4 f^3 h^3}+\frac{675 \left(f'\right)^3}{2 f^3 h^2 r}+\frac{450 h' \left(f'\right)^2}{f^2 h^3 r}+\frac{1035 f'' \left(f'\right)^2}{2 f^3 h^2}+\frac{225 h'' \left(f'\right)^2}{2 f^2 h^3}-\frac{2115 \left(h'\right)^2 \left(f'\right)^2}{8 f^2 h^4}+\frac{90 \left(f'\right)^2}{f^2 h^2 r^2}+\frac{1485 \left(h'\right)^2 f'}{2 f h^4 r}+\frac{270 h' f'}{f h^3 r^2}+\frac{495 h' f'' f'}{f^2 h^3}+\frac{585 h' h'' f'}{2 f h^4}-\frac{225 f^{(3)} f'}{f^2 h^2}-\frac{45 h^{(3)} f'}{f h^3}-\frac{315 \left(h'\right)^3 f'}{f h^5}-\frac{585 f'' f'}{f^2 h^2 r}-\frac{315 h'' f'}{f h^3 r}+\frac{270 f'}{f h r^3}-\frac{90 f'}{f h^2 r^3}+\frac{630 h'}{h^3 r^3}+\frac{855 \left(h'\right)^2 f''}{2 f h^4}+\frac{1170 h' h''}{h^4 r}+\frac{270 f^{(3)}}{f h^2 r}+\frac{90 f^{(4)}}{f h^2}-\frac{315 \left(f''\right)^2}{2 f^2 h^2}-\frac{180 f'' h''}{f h^3}-\frac{270 h' f^{(3)}}{f h^3}-\frac{180 h^{(3)}}{h^3 r}-\frac{630 h' f''}{f h^3 r}-\frac{1260 \left(h'\right)^3}{h^5 r}-\frac{180 f''}{f h^2 r^2}-\frac{270 h'}{h^2 r^3}-\frac{90}{r^4}-\frac{540}{h r^4}+\frac{630}{h^2 r^4}\right) \log \left(\lambda^{2}f\right)\right] \left(\xi -\frac{1}{6}\right)^2,
\end{dmath}
\pagebreak
\end{widetext}
and
\begin{equation}
    \langle\hat{T}^{\varphi}_{\varphi}\rangle_{\rm AHS}=\langle\hat{T}^{\theta}_{\theta}\rangle_{\rm AHS}.
\end{equation}

\bibliographystyle{JHEP}
\bibliography{biblio-semiclassical}

\end{document}